\documentclass{article}

\usepackage{graphicx}

\usepackage{amsmath,latexsym,amssymb}

\usepackage{theorem}

\def\qed{\ifmmode$\Box$\else{\unskip\nobreak\hfil
\penalty50\hskip1em\null\nobreak\hfil$\Box$
\parfillskip=0pt\finalhyphendemerits=0\endgraf}\fi}
\def\endenv{\ifmmode\;\else{\unskip\nobreak\hfil
\penalty50\hskip1em\null\nobreak\hfil\;
\parfillskip=0pt\finalhyphendemerits=0\endgraf}\fi}

\mathchardef\ordinarycolon\mathcode`\:     
\mathcode`\:=\string"8000
\def\vcentcolon{\mathrel{\mathop\ordinarycolon}}
\begingroup \catcode`\:=\active
  \lowercase{\endgroup
  \let :\vcentcolon
  }

\newcommand{\nc}{\newcommand}
\nc{\rnc}{\renewcommand}
\nc{\be}{\begin{equation}}
\newcommand{\ee}{\end{equation}}

\newcommand{\ben}{\begin{enumerate}}
\newcommand{\een}{\end{enumerate}}

\newcommand{\vs}{\vec{\sigma}}

 \newcommand{\cT}{{\cal T}}

 \newcommand{\ha}{\hat{\mu}}
\newcommand{\hb}{\hat{\nu}}

\nc{\bea}{\begin{eqnarray}}
\nc{\eea}{\end{eqnarray}}
\nc{\B}{\cal{B}}
\nc{\lbar}[1]{\overline{#1}}
\nc{\IP}{\operatorname{IP}}
\rnc{\max}{\operatorname{max}}
\nc{\Prob}{\operatorname{Prob}}
\nc{\rank}{\operatorname{rank}}
\nc{\Span}{\operatorname{Span}}
\rnc{\S}{\operatorname{S}}
\nc{\diag}{\operatorname{diag}}
\nc{\sign}[1]{\mbox{sign$(#1)$}}
\nc{\smfrac}[2]{\mbox{$\frac{#1}{#2}$}}
\nc{\Spec}{\operatorname{Spec}}
\nc{\Tr}{\operatorname{Tr}}
\nc{\xor}{\oplus}
\nc{\ox}{\otimes}
\nc{\da}{\dagger}
\nc{\dn}{\downarrow}
\nc{\cA}{{\cal A}}
\nc{\cB}{{\cal B}}
\nc{\cC}{{\cal C}}
\nc{\cD}{{\cal D}}
\nc{\cE}{{\cal E}}
\nc{\cF}{{\cal F}}
\nc{\cG}{{\cal G}}
\nc{\cH}{{\cal H}}
\nc{\cI}{{\cal I}}
\nc{\cJ}{{\cal J}}
\nc{\cK}{{\cal K}}
\nc{\cL}{{\cal L}}
\nc{\cR}{{\cal R}}
\nc{\cS}{{\cal S}}
\nc{\cX}{{\cal X}}
\nc{\cO}{{\cal O}}
\nc{\cU}{{\cal U}}

\nc{\supp}{{\operatorname{supp}}}
\nc{\stab}{{\operatorname{stab}}}
\nc{\ra}{\rightarrow}
\nc{\lra}{\longrightarrow}

\def\S{\Sigma}

\nc{\CC}{{{\mathbb C}}}
\nc{\FF}{{{\mathbb F}}}
\nc{\NN}{{{\mathbb N}}}
\nc{\ZZ}{{{\mathbb Z}}}
\nc{\PP}{{{\mathbb P}}}
\nc{\QQ}{{{\mathbb Q}}}
\nc{\UU}{{{\mathbb U}}}
\nc{\Q}{{{\bar{Q}}}}

\title{\textbf{Non-Computability of Consciousness}}
\author{Daegene Song\thanks{School of Computational Sciences,
Korea Institute for Advanced Study, Seoul 130-722, Korea.; Email: {\tt{dsong@kias.re.kr.}}}}

\begin{document}
\maketitle
\begin{abstract}
With the great success in simulating many intelligent behaviors
using computing devices, there has been an ongoing debate whether
all conscious activities are computational processes. In this
paper, the answer to this question is shown to be no.  A certain
phenomenon of consciousness is demonstrated to be fully
represented as a computational process using a quantum computer.
Based on the computability criterion discussed with Turing
machines, the model constructed is shown to necessarily involve a
non-computable element. The concept that this is solely a quantum
effect and does not work for a classical case is also discussed.

\end{abstract}
\section{Introduction.}
Research in the field of artificial intelligence, which attempts
to imitate and simulate intelligent activities using a machine,
has blossomed along with the development of information technology
\cite{AI}.  Because the study of artificial intelligence has
provided many insights into intelligent behaviors such as pattern
recognition, decision theory, etc., there is a question whether
consciousness or self-awareness could emerge out of a
computational system, a view termed as strong artificial
intelligence. This question can be rephrased and stated as
follows: {\it{Are all conscious activities computational
processes?.}} In this paper, the answer to this question is shown
to be no.

In order to examine the computability of a physical phenomenon,
the phenomenon should first be represented as a computational
model; subsequently, the computability of this particular model
can be examined.  The physical phenomenon can then be claimed to
be computable or not based on this examination. A similar approach
will be taken in order to examine the computability of
consciousness. Because consciousness is a phenomenon experienced
by an observer, representation of consciousness as a computational
process will be attempted and its computability will be examined.
Although traditional approaches for studying consciousness have
included neuroscience \cite{koch} or neural network modeling
\cite{tononi,harvey}, it is demonstrated that a quantum system to
be presented below, it necessarily involves a conscious, as
opposed to a physical, activity of an observer observing the
unitary dynamics of a quantum state.  Based on this observation, a
particular quantum computer can be built such that it yields a
computational model involving consciousness. Using logic similar
to that in Turing's haling problem, it can be shown that this
computational model necessarily runs into a contradiction. As a
result, this effectively provides a counter-example to the
assumption that all conscious activities are computational
processes.

In this paper, it is not claimed that all conscious activities can
be constructed, using a quantum computer, nor that they are
quantum mechanical. Instead, it will be argued that the quantum
system to be presented necessarily involves a certain conscious
activity and that quantum theory provides a full description of
this particular conscious activity.  This argument will be used to
build a quantum computing machine such that it suffices to provide
a counter-example. A single counter-example is sufficient to prove
the assumption is incorrect.

\begin{figure}
\begin{center}
{\includegraphics[scale=.6]{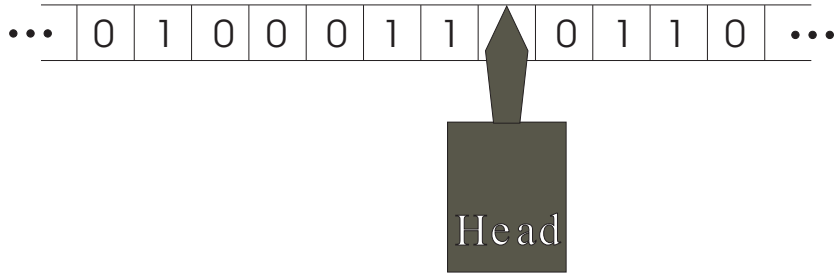}}
\end{center}
\caption{Turing machine. A Turing machine is an abstract model of
a computing system consisting of internal states, a tape
containing symbols in each cell and a head that reads and writes
the symbol. Evolution in time of the Turing machine is described
by $(I,a)\rightarrow (I^{\prime},a^{\prime},d)$ where $I$ is the
internal state, and $a$ is a symbol written on the tape.  At the
$i$th cell, i.e., the head's position, the head reads the symbol
$a$, and, with the instruction $I$, it writes a new symbol,
$a^{\prime}$, and moves either one cell to the left ($d=-1$) or to
the right ($d=+1$) with an updated internal state, $I^{\prime}$.
The initiation and termination of computation are indicated by
internal states, $h_0$ and $h_1$, respectively. }
\label{Turing}\end{figure}

\section{Computability and Turing machine.}
In order to discuss computability of consciousness, let us first
consider what it means to be computable.  This can be done using
the notion of Turing machines. A Turing machine, denoted as
${\rm{TM}}$, is a theoretical model of dynamic computing system
configured with an internal state, a tape containing a symbol in
each cell and identification of the position of the head on the
tape (see Fig. \ref{Turing}).  The time evolution of the
${\rm{TM}}$ is described by $(I,a)\rightarrow
(I^{\prime},a^{\prime},d)$ where $I$ is the internal state, $a$ a
symbol in the cell of the tape, and $d=\pm 1$. Therefore, at the
$i$th cell, the head reads the symbol, $a$. Then, given the
current internal state, $I$, which provides an instruction, the
new symbol, $a^{\prime} = I(a)$, is written, and the internal
state is updated to $I^\prime$ and the head moves either one cell
to the right ($d=+1$) or one cell to the left ($d=-1$), i.e., to
the ($i+d$)th cell.

Among the functions of the internal state, is the indication of
initiation and termination of computation. Initially, this
particular state is set to $h_0$, indicating the initiation of the
computation. After the computation is completed, the state is set
to $h_1$ and the machine ceases its activity. The output of the
computation corresponds to the symbol in the cell where the head
is located when the machine halts.  For a given input, $i$, the
${\rm{TM}}$ runs, following the time evolution described above,
and either (A) produces an outcome, $f={\rm{TM}}(i)$, and halts
with the internal state set to $h_1$ or (B) loops forever and the
internal state never reaches $h_1$.  A system is called computable
if it corresponds to a ${\rm{TM}}$ such that it follows either (A)
or (B) for a given input $i$, and is called non-computable
otherwise.

\begin{figure}
\begin{center}
{\includegraphics[scale=.5]{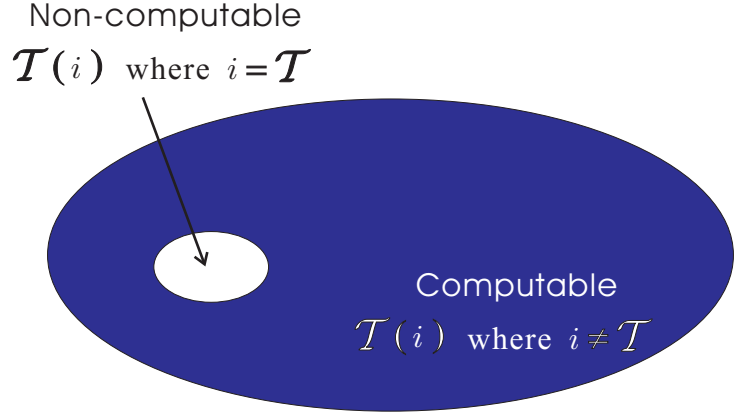}}
\end{center}
\caption{Computability and Non-computability. $\cT$ is defined to
have the property corresponding to a Turing machine that either
halts or not unless it is given an input of $\cT$ itself. The
halting problem can be defined in association with $\cT$, such
that it necessarily is non-computable.  Similarly, quantum theory
allows consciousness to be represented as a computational process
in terms of $\cT$, such that it would necessarily consist of a
non-computable element when the input is $\cT$ itself.
 } \label{computable}\end{figure}

The issue of computability is considered in the following setup:
suppose $\cT$ is defined to have the following two properties:
\begin{enumerate}
\item Computable: $\cT (i)$ when $i\neq \cT$    \item
Non-computable: $\cT (i)$ when  $i = \cT$
\end{enumerate}
That is, $\cT$ corresponds to a ${\rm{TM}}$ that follows either
(A) or (B) except when the input is $\cT$, i.e., the description
of $\cT$, itself (see Fig. \ref{computable}). In the following
approach, the manner in which the computational model involving
consciousness may be defined through $\cT$ will be shown using a
quantum computing machine, such that the two conditions regarding
computability are satisfied, i.e. necessarily containing
non-computability.

\section{The halting problem.}
Before proceeding with the discussion of consciousness, it is
instructive to review Turing's halting problem \cite{turing} and
to examine its use of a property similar to that seen in $\cT$,
such that the problem was shown to be non-computable. The
situation for the halting problem is as follows: for some
${\rm{TM}}$s, an outcome indicated by the halt internal state
$h_0\rightarrow h_1$, is computed, while for some other
${\rm{TM}}$s, with a given input, the computation loops forever
indicated by a constant internal state, $h_0$.  Turing's halting
problem asks if there is a ${\rm{TM}}$ that can distinguish
between the two types given an arbitrary ${\rm{TM}}$ and an input.
Suppose there is such a ${\rm{TM}}$ that performs a calculation
given the description of ${\rm{TM}}$ and $i$ such that it is able
to determine if ${\rm{TM}}$ halts or not. This assumption then
makes it possible to construct a particular ${\rm{TM}}$,
${\rm{TM_H}}$, such that the machine does not halt, for an input
${\rm{TM}}$, if and only if ${\rm{TM(TM)}}$ halts. However, a
contradiction follows for ${\rm{TM_H}}$ when the input is
${\rm{TM_H}}$ itself, because ${\rm{TM_H (TM_H)}}$ does not halt,
if and only if, ${\rm{TM_H (TM_H)}}$ halts.

Therefore, by identifying the ${\rm{TM_H}}$ associated with the
hypothetical ${\rm{TM}}$ that could decide if an arbitrary
${\rm{TM}}$ would halt on a given input, it is possible to show
that the ${\rm{TM_H}}$ contains an element that neither halts
after completing the computation nor loops forever as should a
valid ${\rm{TM}}$. The constructed ${\rm{TM_H}}$ has the same
property as $\cT$ in Fig. \ref{computable}, i.e., it is computable
except when the input is ${\rm{TM_H}}$ itself.

\section{Conscious activity in quantum system.}
In order to represent a phenomenon of consciousness as a
computational model, the manner in which a conscious activity is
involved in a quantum system is first discussed.  This will be
conducted using the notation of a qubit, a two-level quantum
system. A qubit in a density matrix form is written as
$|\psi\rangle\langle \psi| = \frac{1}{2}({\bf {1}}+ {\bf{\ha}}
\cdot \vs )$ where $ {\bf {\ha}}=(\mu_x,\mu_y,\mu_z)=(\sin\theta
\cos\phi,\sin\theta\sin\phi,\cos\theta )$
  and  $\vs = (\sigma_x,\sigma_y,\sigma_z)$ with
$\sigma_x = |0\rangle\langle 1| + |1\rangle\langle 0|$, $\sigma_y
= -i|0\rangle\langle 1| + i|1\rangle\langle 0|$, and $\sigma_z =
|0\rangle\langle 0| - |1\rangle\langle 1|$. Therefore a qubit,
$|\psi\rangle\langle \psi|$, can be represented as a unit vector
$\ha = (\mu_x,\mu_y,\mu_z)$ pointing in $(\theta,\phi)$ of a
sphere with $0\leq \theta \leq \pi , 0\leq \phi \leq 2\pi$.   In
quantum theory, there is another important variable called an
observable. For a single qubit, an observable can also be written
as a unit vector, $\hb=( \nu_x,\nu_y,\nu_z )=(\sin\vartheta
\cos\varphi,\sin\vartheta\sin\varphi,\cos\vartheta )$, pointing
$(\vartheta,\varphi)$ direction in a sphere. Therefore if one is
to make a measurement in $(\vartheta,\varphi)$ direction,
 the observable would be $\hb \cdot \vs$.  Representing a qubit
and an observable as unit vectors in the Bloch sphere will make
their visualization easier which will be helpful in the following
 discussions.

Let us consider one particular phenomenon, denoted as ${\bf{P1}}$,
and described as follows: {\it{an observer observes the unitary
evolution of a qubit, $\ha$, with respect to the observable,
$\hb$.}} The observer is observing the evolution of $\ha$
indirectly and a measurement can be followed in order to confirm
the evolution. When a measurement on $\ha$, with the observable,
$\hb$, is made, it yields a real eigenvalue that can be directly
observed by the observer. Before discussing the description of the
phenomenon, ${\bf{P1}}$, using the dynamics of quantum theory, it
is necessary to illustrate why the phenomenon, ${\bf{P1}}$,
necessarily involves a conscious activity of the observer. An
observable serves as a coordinate or a reference frame when the
measurement is made on a given state vector \cite{peres}. This
concept is easier to visualize with two unit vectors, $\ha$ and
$\hb$. The unit vector representing an observable, i.e., $\hb$, is
serving the role of a coordinate for the unit vector representing
a qubit, $\ha$. Because the measurement is performed by an
observer, the observable is considered to be a coordinate or a
reference frame of the observer, for a given qubit $\ha$.

However, in quantum theory, observables, being a reference frame
of the observer, are fundamentally different from reference frames
in classical physics.  In quantum theory, the state vectors have
representation in, and evolve in, the Hilbert space, a complex
vector space.  This description was invented in order to correctly
predict the outcome of measurement performed on a state vector
which yields a real eigenvalue outcome. Not only do state vectors
reside and evolve in the Hilbert space, so do the observables.
Because the observables correspond to the reference frame of the
observer and they exist in the complex Hilbert space, it must be
concluded that, unlike reference frames in classical physics,
quantum observables correspond to an observer's reference frame in
thought. That is, an observable should be considered to represent
the conscious status of an observer while observing a given state
vector.  This argument explains why the phenomenon, ${\bf{P1}}$,
necessarily involves a conscious activity.

The qubit in ${\bf{P1}}$ can be any 2-level quantum system, for
example, a spin 1/2-particle or any quantum system in 2-levels,
etc. However, it is not necessary to specify all properties of the
physical system other than $\ha$, because $\ha$ is a pure state
and is disentangled from the state that represents other
properties of that quantum system. Therefore, as far as the
phenomenon, ${\bf{P1}}$, is concerned, $\ha$ provides a full
description of the physical system. The same logic applies to
$\hb$ as well. The vector, $\hb$, is not entangled with vectors
representing other observables. Therefore, $\hb$ must provide a
full description of the conscious status of the observer in
phenomenon, ${\bf{P1}}$. That is, similarly to the case with
$\ha$, it is not necessary to be concerned with other conscious
activities of the observer because $\hb$ is disentangled from
them. Therefore, ${\bf{P1}}$ not only necessarily involves a
conscious activity of the observer, $\hb$ gives a full description
of the conscious activity as far as the phenomenon, ${\bf{P1}}$,
is concerned.

Quantum theory provides two approaches in describing the natural
phenomenon ${\bf{P1}}$.  Given $\ha$ and $\hb$, the first is by
applying a unitary operation to the qubit with $\ha^{\prime} = U
\ha U^{\dagger}$ where a measurement would yield the expectation
value of $ \hb \cdot \ha^{\prime}$. The second is by applying a
unitary operation to the observable as $\hb^{\prime} = U^{\dagger}
\hb U$ and a measurement would yield $\hb^{\prime}\cdot \ha$. That
is, quantum theory insists that, in order to have an observer
observe the unitary transformation of $\ha$ with respect to $\hb$,
either a unitary transformation is applied to the qubit, i.e., the
first approach, or the observer's reference frame $\hb$ is
changed, i.e., the second approach.  The first approach is called
the Schr\"odinger picture and the second corresponds to the
Heisenberg picture.

In the second approach, it was the observable that went through a
unitary transformation which should describe the same phenomenon,
${\bf{P1}}$, as the first approach. Because the evolution of
observables through unitary transformations are performed in the
Hilbert space and the observable is the observer's conscious
status in ${\bf{P1}}$, an observable that is being changed must
correspond to a conscious activity of an observer. However, while
the observer's conscious status is being changed, the observer is
not observing the observable but the state vector, $\ha$.
Therefore, this approach also yields the description of the
natural phenomenon, ${\bf{P1}}$, just as in the first approach.


\section{Conscious activity in quantum computing process.}
So far, it has been argued that the phenomenon, ${\bf{P1}}$,
necessarily involves an observer's conscious activity and quantum
theory provides a full description of the conscious status of the
observer regarding ${\bf{P1}}$. Based on these observations, a
quantum computational model is to be constructed such that it
represents a phenomenon involving a conscious activity and its
computability will be examined.   In particular, $\cT$ will be
defined in terms of this computational model and the computability
for a given input, $i$, will be examined. In the next section, it
will be argued that when the input is $\cT$ itself, it represents
consciousness and will be proven to be non-computable similarly to
the halting problem.

Let us review basic elements of quantum computation by following
the discussion in \cite{benioff,deutsch}. The particular class of
quantum computers to be considered is assumed to perform a
computation on an input of a single qubit, i.e., a unit vector in
the Bloch sphere, $\ha_s$, in which the subscript, $s$, is placed
in order to distinguish it from the halt qubit to be defined
shortly. In this particular class, the computation is assumed to
be conducted through a unitary process on a given single input
qubit, a rotation about the $y$-axis by $\delta$, i.e., $U_y
\equiv \cos\frac{\delta}{2}|0\rangle\langle 0|
-\sin\frac{\delta}{2}|0\rangle\langle 1| +\sin\frac{\delta}{2}
|1\rangle\langle 0| + \cos\frac{\delta}{2}|1\rangle\langle 1|$.
Among the components of the classical ${\rm{TM}}$, the head exists
which reads each cell on the tape (see Fig. \ref{Turing}). The
head in the classical ${\rm{TM}}$ may correspond to the
observables in the quantum computer. As demonstrated earlier, for
a single qubit, the observable can also be written as a unit
vector in the Bloch sphere, which will be denoted as $\hb_s$.

As suggested in \cite{deutsch}, in addition to the system input
qubit, an additional qubit is placed which indicates if the
computation on the system qubit has successfully ended by
$0\rightarrow 1$ after a valid computation on the single system
qubit which remains $0$ otherwise. This is equivalent to the
classical ${\rm{TM}}$ in which its internal state indicates if the
machine completed its computation by $h_0 \rightarrow h_1$. The
halt qubit is set to point at the $z$-direction, i.e. $\ha_h =
(0,0,1)$.  The corresponding observable, $\hb_h=(0,0,1)$, also set
to point at the $z$-direction, initially. Therefore, the quantum
computer constructed for a given input $\ha_s$, is a closed system
consisting of $\ha_s,\hb_s,\ha_h$, and $\hb_h$. Because there is
freedom to set the observable, it can be used to identify one
particular quantum computer which works on a given input, $\ha_s$.
Among the infinitely many choices of $\hb_s$, assume that one
particular quantum computer exists with the observable, $\hb_s =
(0,0,1)$. Because the unitary evolution will be $U_y$ only, the
initial observable fully characterizes this particular quantum
computer and it will be defined as $\cT$.

The quantum model constructed operates on a single qubit, and,
only a single operation, i.e., $U_y$, is considered.  Therefore,
there is no need to specify any internal state that yields an
instruction because there is only one operation.  The only
internal state needed is the indication of initiation and
termination of the computation that is represented with the halt
qubit. Moreover, indication of the position of the head is
unnecessary because there is only one qubit, which corresponds to
a tape with a single cell. Therefore, the quantum computer
constructed corresponds to a very simple case of a quantum
mechanical ${\rm{TM}}$.

One particular phenomenon, denoted as ${\bf{P2}}$, is considered
as follows: {\it{an observer observes a rotation of the input,
$\ha_s$, about the $y$-axis by $\delta$, with respect to
$\hb_s$.}} As in ${\bf{P1}}$, the observer is observing the
rotation indirectly and a measurement on $\ha_s$ with the
observable, $\hb_s$, can be followed to confirm the evolution.
Note that ${\bf{P2}}$ is almost identical to the phenomenon,
${\bf{P1}}$, except the unitary operation is specified as $U_y$.
Therefore, similarly to the case with ${\bf{P1}}$, the phenomenon,
${\bf{P2}}$, necessarily involves a conscious activity of the
observer, represented as $\hb_s$. Moreover, as discussed with the
instance of ${\bf{P1}}$, $\hb_s$ provides a full description of
the conscious status of the observer in reference to ${\bf{P2}}$.
In the following, it will be established that the quantum computer
constructed, $\cT$, represents ${\bf{P2}}$ as a computational
model and is computable, therefore indicating that the phenomenon,
${\bf{P2}}$, is computable.

As discussed earlier, the quantum theory provides two approaches
for the evolution in time of a quantum system. Therefore, because
$\cT$,  the quantum computer constructed, is a quantum system, it
should also evolve in both approaches. The evolution in time of
$\cT$ with an initial input state, $i=\ha_s=(0,0,1)$, will be
examined. The first approach, i.e., the Schr\"odinger picture, is
considered as follows: the unitary operation $U_y$ transforms the
input as $\ha_s \rightarrow U_y \ha_s U_y^{\dagger}$ and the halt
qubit $\ha_h \equiv (0,0,1)$ halts by transforming into $-\ha_h$.
In the second approach, i.e., the Heisenberg picture, it is the
observable that evolves. Therefore, $U_y^{\dagger}$ transforms the
vector representing the observable $\hb_s$ into
$U_y^{\dagger}\hb_s U_y$ and the observable for the halt qubit
$\hb_h\equiv (0,0,1)$ is transformed into $-\hb_h$. Therefore, in
the second approach, the observer's conscious status $\hb_s$ is
being changed while the observer observes $\ha_s$. This should
yield the same observation as the first approach. It is noted that
the expectation value of $(U_y^{\dagger} \hb_s U_y)\cdot \ha_s$
for the second approach is equal to the expectation value in the
first approach, $\hb_s \cdot (U_y \ha_s U_y^{\dagger})$.
Therefore, both the first and the second computational processes
ultimately describe the phenomenon, ${\bf{P2}}$, by correctly
producing an outcome described in ${\bf{P2}}$.

Initially, it was discussed that a system is stated to be
computable when it satisfies one of two criteria, i.e. either (A)
it halts after completion of a valid computation or (B) it loops
forever without halting.  $\cT$ was shown to yield the description
of ${\bf{P2}}$, with a given input $\ha_s$, by following both
pictures in quantum theory, i.e., both approaches yielded the
outcome by which $\ha_s$ rotated about the $y$-axis by $\delta$,
with respect to $\hb_s$, and halted. Therefore, the phenomenon,
${\bf{P2}}$, can be claimed to be computable because its
computational representation, $\cT$, with the input $\ha_s$, was
shown to be computable by satisfying the criterion (A).

\section{Counter-Example to the Assumption.}
In case of the Heisenberg picture description of ${\bf{P2}}$, as
well as of ${\bf{P1}}$, it was discussed that the observer is in
the conscious status undergoing change, $\hb_s$, and observes
$\ha_s$. This was shown to yield the phenomenon of ${\bf{P2}}$,
i.e., the observer observing the rotation of $\ha_s$. A slightly
different case can be considered. While the observer is in the
conscious status, $\hb_s$, that is being changed, the observer
observes $\hb_s$ rather than $\ha_s$. This is a peculiar aspect of
consciousness--observing one's own conscious status--that is not
observed in other measurement experiences, for example, in
classical dynamics. This phenomenon can be stated as follows and
denoted as ${\bf{P3}}$: {\it{an observer observes a rotation of
the input, $\hb_s$, about the $y$-axis by $\delta$, with respect
to $\hb_s$}}. Therefore, in ${\bf{P3}}$ which describes
consciousness of the observer, $\hb_s$ is serving the role of a
state vector, because it is being observed, and an observable,
because it is serving as the reference frame of the observer.
Unlike the cases of ${\bf{P1}}$ and ${\bf{P2}}$, the measurement
confirmation is not needed for ${\bf{P3}}$. While the conscious
status, $\hb_s$, is evolving, the observer is not observing
$\ha_s$ but $\hb_s$. No measurement is needed in order to confirm
the evolution of $\hb_s$ because the observer is already
experiencing it as consciousness.

In the previous section, it was demonstrated that $\cT$, with an
input $\ha_s$, provides a computational model for describing the
phenomenon of ${\bf{P2}}$ and was shown to be computable. Because
${\bf{P3}}$ is exactly the same as ${\bf{P2}}$ except the input
has changed to vector, $\hb_s$, from $\ha_s$, it follows that
$\cT$, with an input, $\hb_s$, must correspond to a computational
model representing the phenomenon, ${\bf{P3}}$ (see Table
\ref{table}). The observable, $\hb_s=(0,0,1)$, fully characterizes
$\cT$. Therefore, $\cT$ with an input, $\hb_s$, can also be stated
as $\cT$ with an input of the description of $\cT$, or simply as
$\cT$ with an input, $\cT$. In the following, the computability of
$\cT$ for a given input of $\cT$, which represents the phenomenon,
${\bf{P3}}$, as a computational model, is to be examined.

\begin{table}
\begin{center}
\begin{tabular}{|l |l |} \hline
  {\bf{Computational Model}} &  {\bf{Phenomenon}} \\
  \hline \hline
  ($\cT$,$i=\ha_s$)  &  ${\bf{P2}}$: Observer observes the rotation of $\ha_s$ with respect to
  $\hb_s$.  \\
  \hline
 ($\cT$,$i=\hb_s$)&  ${\bf{P3}}$: Observer observes the rotation of $\hb_s$ with respect to $\hb_s$. \\
 \hline
 \end{tabular}
 \end{center}
 \caption{Analogy between the computational model, $\cT$, and phenomena ${\bf{P2}}$ and ${\bf{P3}}$.
 If the phenomenon, ${\bf{P2}}$, can be represented as a computational model, $\cT$, with an input, $\ha_s$, then
 $\cT$ with an input, $\hb_s$, should correspond to a computational model for the phenomenon, ${\bf{P3}}$.    }
\label{table}\end{table}

As established previously, quantum theory provides two approaches
to the evolution in time of $\cT$ for the input, $\hb_s$, because
it is a quantum system where $\hb_s$ corresponds to both a state
and an observable. In the first approach, it is the input system
that evolves. Since the input is $\hb_s$, the evolution is as
follows, $\hb_s \rightarrow \hb_s^{\prime} =
(\sin\delta,0,\cos\delta)$, while the halt qubit is transformed as
$\ha_h\rightarrow -\ha_h$. Quantum theory provides a second
approach where the same vector, being an observable, is
transformed as $\hb_s \rightarrow \hb_s^{\prime\prime} =
(-\sin\delta,0,\cos\delta)$, while the observable for the halt
qubit is transformed as $\hb_h\rightarrow -\hb_h$. It is noted
that $\hb_s^{\prime} \neq \hb_s^{\prime\prime}$ unless $\delta =
k\pi$ where $k=0,1,2,...$.

Let us now discuss the computability of $\cT(i)$ where $i=\cT$. In
order for $\cT(\cT)$ to be computable, it has to follow either the
computability criterion (A) or (B).  Since $\cT$ halted on both
approaches, i.e., $\ha_h \rightarrow -\ha_h$, with respect to
$\hb_h$, in both pictures, it must follow (A) rather than (B) in
order to be computable.  In order to satisfy (A), the halt qubit
of $\cT$ must have halted accompanied by a valid computation,
i.e., both approaches should yield the same outcome predicted in
${\bf{P3}}$. However, the two approaches yielded two generally
different outcomes for a single input vector, $\hb_s$. The second
approach did not yield the outcome described in the phenomenon,
${\bf{P3}}$, because the vector is rotated by $-\delta$.
Therefore, this results in a contradiction because $\cT$ halted on
the invalid computation. The contradiction is noted to result from
a peculiar property of consciousness in which $\hb_s$ is serving
as a reference frame of the observer and as a system to be
observed.

The assumption states that all conscious activities are
computational processes.  Because $\cT(i)$, with $i=\hb_s$, being
a computational model of the phenomenon ${\bf{P3}}$, is a closed
and independent system, this must satisfy the assumption. However,
it was shown that $\cT$, with a given input $\hb_s$, is not
computable. That is, a particular conscious activity of an
observer observing the change of an observable, as described in
${\bf{P3}}$, is not computable. Therefore, this leads to a
conclusion that the assumption is incorrect, because it suffices
to have a single counter-example to invalidate the assumption.

Perhaps, by considering a larger system that includes the qubit,
the contradiction may be removed and may yield the result that
consciousness is always a computational process.  This is commonly
seen in thermodynamics in which a subsystem violates the second
law but this violation is always removed when the total system is
considered.  However, this kind of argument would not work because
the evolution considered in ${\bf{P2}}$ and ${\bf{P3}}$ are for
pure states.   Any attachment of ancilla to $\cT$ and their
interaction with the system qubit would cause entanglement and
this will not properly represent the physical phenomena
${\bf{P2}}$ and ${\bf{P3}}$.

\section{Discussion.}
The above argument applies only as a quantum effect.  The
classical ${\rm{TM}}$ cannot define consciousness using the same
technique.  As discussed, a reference frame of quantum measurement
was represented in complex Hilbert space which led to the
conclusion that it must correspond to the observer's conscious
status. A classical measurement yields an outcome in terms of the
difference between the object and the reference frame of an
observer, and, unlike consciousness, the observer cannot observe
the dynamics of its own reference frame alone. Therefore, the same
argument used with the quantum computing machine involving
conscious activities cannot be used in a classical case.

In \cite{penrose}, Penrose discussed that a non-computable aspect
in consciousness may exist at the fundamental level as described
in G\"odel's incompleteness theorem. Including Turing's halting
problem, there have been a number of mathematical examples showing
undecidability in G\"odel's theorem. In this paper, it was
demonstrated that, as in Penrose's suggestion, consciousness is a
physical, i.e., rather than mathematical, example of G\"odel-type
proofs.

\end{document}